\documentclass{rspublic}
\usepackage{graphics}

\begin{document}

\newcommand{\one}{\ensuremath{\hbox{$\mit I$\kern-.3em$\mit I$}}}
\renewcommand{\one}{\ensuremath{\hbox{$\mathrm I$\kern-.6em$\mit 1$}}}
\renewcommand{\one}{\ensuremath{\hbox{$\mathrm I$\kern-.6em$\mathrm 1$}}}
\newcommand{\ad}{\ensuremath{a^\dagger}}
\newcommand{\bd}{\ensuremath{b^\dagger}}
\newcommand{\cd}{\ensuremath{c^\dagger}}
\newcommand{\dd}{\ensuremath{d^\dagger}}
\newcommand{\ed}{\ensuremath{e^\dagger}}
\newcommand{\ket}[1]{\ensuremath{|{#1}\rangle}}
\newcommand{\bra}[1]{\ensuremath{\langle{#1}|}}
\newcommand{\braket}[1]{\ensuremath{\langle{#1}\rangle}}

\title{Quantum computation with cold bosonic atoms in an optical lattice}

\author[J.~J. Garc\'{\i}a-Ripoll \& J.~I. Cirac]%
  {Juan Jos\'e Garc\'{\i}a-Ripoll \& Juan Ignacio Cirac}
\affiliation{Max-Planck-Institut f\"ur Quantenoptik,
  Hans-Kopfermann-Str. 1\\ Garching D-85748, Germany.}
\label{firstpage}
\maketitle
\begin{abstract}{Quantum computing, adiabatic, cold atoms, optical lattice.}
  We analyse an implementation of a quantum computer using bosonic atoms in an
  optical lattice. We show that, even though the number of atoms per site and
  the tunneling rate between neighbouring sites is unknown, one may perform a
  universal set of gates by means of adiabatic passage.
\end{abstract}


\section{Introduction}

Some tasks in Quantum Information require the implementation of quantum gates
with a very high fidelity (A. M. Steane 2002; E. Knill 1998; D. Aharonov 1999).
This implies that all parameters describing the physical system on which the
computer is implemented have to be controlled with a very high precision,
something which it is very hard to achieve in practice.

For example one can imagine an implementation in which qubits are stored in
atoms and are manipulated using Raman transitions. It may happen that the
relative phase of the lasers driving a Raman transition can be controlled very
precisely, whereas the corresponding Rabi frequency $\Omega$ has a larger
uncertainty $\Delta\Omega$.  If we denote by $T$ the time required to execute a
local gate (of the order of $\Omega^{-1}$), then a high gate fidelity requires
$T\Delta\Omega\ll 1$ (equivalently, $\Delta\Omega/\Omega\ll 1$), which may very
hard to achieve, at least to reach the above mentioned threshold.

In this paper we analyse an implementation of quantum computing using atoms
confined in optical lattices. These systems have interesting features for
quantum computing. Namely, a large number of atoms can be trapped in the
lattice at a very low temperature, which provides a large number of qubits.
Also, neutral atoms interact weakly with the environment, which leads to a
relatively slow decoherence. However, the same setups pose also important
experimental challenges, such as being able to load the lattice with one atom
per site or being able to measure the interaction and tunneling constants of
these systems with high accuracy. These obstacles, together with the
uncertainties in the atom-laser interaction, must be overcome to implement
current proposals for quantum computing with neutral atoms (D.  Jaksch 1999; R.
Ionicioiu 2002).

In this paper we analyse a way that solves the above mentioned problems and
show how to achieve a very high gate fidelity even when most of the parameters
describing the atomic ensemble (number of atoms per lattice site, tunneling
rate, Rabi frequencies, etc) cannot be adjusted to precise values, and even
have uncertainties of the order of the parameters themselves.

Our method combines the technique of adiabatic passage with ideas of quantum
control theory. The use of adiabatic passage to implement quantum gates is not
a new idea, and indeed several methods based on Berry phases have been proposed
recently (P. Zanardi 1999; J.  Pachos 2000; J. A.  Jones 1999; G. Falci 2000;
L.-M. Duan 2001).  Furthermore, adiabatic passage techniques have been
suggested as a way of implementing a universal set of holonomies (L.-M. Duan
2001), i.e. quantum gates which are carried out by varying certain parameters
and whose outcome only depends on geometrical properties of the paths in
parameter space (P. Zanardi 1999; J.  Pachos 2000).  However, all these
proposals are based on the existence of holonomies in the system, which in turn
implies a huge degeneracy in the system. This will not be the case in our
scheme.

The outline of the paper is as follows. In \S\ref{sec:lattice} we introduce the
requirements for quantum computing and show which tools are available in
current experiments with cold atoms in optical lattices. We will demonstrate
that, due to imperfections in the loading of the lattice, the Hamiltonian of
the system is not known with enough accuracy to perform quantum computing in
a `traditional' way. In \S\ref{sec:adiabatic} we develop a technique to
circumvent our ignorance about the Hamiltonian. Performing adiabatic passage
with the different parameters of our problem, we show how to produce a
universal set of gates (Hadamard, phase, and CNOT). In \S\ref{sec:errors} we
quantify the errors of our proposal, studying the influence of the speed of the
adiabatic process, and of other imperfections. In \S\ref{sec:conclusions} we
summarize our results and offer some conclusions.


\section{Cold bosonic atoms in optical lattices}
\label{sec:lattice}

We will consider a set of bosonic atoms confined in a periodic optical lattice
at sufficiently low temperature such that only the first Bloch band is
occupied. The atoms have two relevant internal (ground) levels, $\ket{a}$ and
$\ket{b}$, and we wish to use this degree of freedom to store the qubit.  This
set-up has been studied in D. Jaksch (1999) where it has been shown how single
quantum gates can be realized using lasers and two--qubit gates by displacing
the atoms that are in a particular internal state to the next neighbour
location. The basic ingredients of such a proposal have been recently realized
experimentally (I. Bloch 2002, personal communication). However, in this and
all other schemes so far (E. Charron 2002; K. Eckert 2002) it is assumed that
there is a single atom per lattice site since otherwise even the concept of
qubit is no longer valid.  In present experiments, in which the optical lattice
is loaded with a Bose-Einstein condensate (D. Jaksch 1998; M.  Greiner 2002),
this only approximately true, since zero temperature is required and the number
of atoms must be identical to the number of lattice sites.

\subsection{Requirements for computation}

The uncertainty of the number of atoms per lattice site poses severe problems.
Having $n_i$ atoms on the $i$-th cell, the configuration of this lattice size
will be given by a combination of $n_i+1$ possible states
\begin{equation}
  \label{all-states}
  {\cal H}_i = \mathrm{lin}\left\{
  \frac{1}{\sqrt{n_i!(n_i-k)!}}{\bd_i}^k{\ad_i}^{n_i-k}\ket{vac}_i,
  \quad k=0\ldots n_i\right\}.
\end{equation}
To do quantum computing with $m$ qubits, we must find a $2^m$-dimensional
subspace ${\cal H}_c \subset {\cal H}_1\otimes\cdots\otimes {\cal H}_m$, which
is energetically separated from the rest, so that once we set our computer in a
superposition of qubits \ket{0} and \ket{1}, it does not leave this subspace. A
second, and stronger requirement is that our computation space must be an
eigenspace of our Hamiltonian, with the same eigenvalue
\begin{equation}
  \label{requirement-1}
  H (\ket{z_1}\otimes\cdots\otimes\ket{z_n}) = 0, \quad
  \forall z_i \in \mathbb{Z}_2,
\end{equation}
which we assume 0. Otherwise the trivial evolution of our system would spoil
the quantum computation by introducing uncontrollable, unknown
phases.

Right from the beginning we forsee several difficulties. First, for arbitrary
interactions, the states with different occupation numbers (\ref{all-states})
will be regularly spaced and we will not be able to select our qubits.
Furthermore, even if we customize the interactions between bosons, given that
we basically ignore the number of atoms per site, $n_i$, a basic requisite of
our scheme will be to show that our procedure works independently of the
occupation numbers. Both problems cannot be solved in general. We will rather
have to impose some restrictions on our physical system, and this is the
purpose of the following subsections.

\subsection{Definition of qubit}

A crucial assumption which is suggested by the requirement
(\ref{requirement-1}), is to impose that the atoms in internal state $\ket{a}$
do not interact and do not hop to neighbouring sites\footnote{This may be
  achieved by tuning the scattering lengths and the optical lattice}. In the
absence of external fields, the Hamiltonian describing our system is
\begin{equation}
\label{H-orig}
 H = \sum_k \left[- J_{b,k} (\bd_{k+1}b_k + \bd_kb_{k+1})
 + \frac{1}{2}U_{bb}\, \bd_k\bd_kb_kb_k\right].
\end{equation}
Here $U_{bb}$ and $J_{b,k}$ describe the interactions between and the
tunneling of atoms in state $\ket{b}$. We will assume that $J_{b,k}$ can be
set to zero and increased by adjusting the intensities of the lasers which
create the optical lattice.

For us a qubit will be formed by an aggregate of {\em at least one} atom per
lattice site and the qubit basis will be formed by the states with at most one
atom excited to the state $b$. More precisely, our computation will be
performed in the space
\begin{equation}
  \label{qubit-space}
  {\cal H}_c(\vec{n}_i) = \left\{\ket{z_1\ldots z_M}=
    \prod_{i=1}^{M}\frac{1}{\sqrt{n_i!}}\,
    {\bd_i}^{z_i}{\ad_i}^{n_i-z_i}\ket{vac}
    : \forall z_i \in \mathbb{Z}_2\right\}.
\end{equation}
It is easy to check that for $J_b=0$, all our qubit states $\ket{b_1\ldots
  b_M}$ form degenerate linear eigenspace of our Hamiltonian (\ref{H-orig}),
which is separated by an energy gap of $U_{bb}$ from any other configuration.

Definition (\ref{qubit-space}) fulfills some of the requirements for quantum
computing. However, ${\cal H}_c$ does depend on the occupation number of the
lattice, while a general state will be an incoherent superposition of different
occupation numbers. It remains to show that we are able to produce quantum
gates which are insensitive to the numbers $n_i$. More precisely, if we design
a protocol to produce the gate $U_{id}$, and this protocol is implemented by
the unitary operation $U_{real}$, we must prove that within the required
accuracy
\begin{equation}
  \label{requirement-2}
  \left.U_{real}\right|_{{\cal H}_c(\vec{n}_i)} \simeq
  e^{i\phi(\vec{n}_i)}\,U_{id}.
\end{equation}
The phases $\phi(\vec{n}_i)$ are irrelevant, since they are common to each of
the possible computation spaces and final measurements will project our state
to one of the subspaces ${\cal H}_c$.

\subsection{Available tools}

The quantum gates will be realized using lasers, switching the tunneling
between neighbouring sites, and using the atom--atom interaction. We will now
show how these elements introduce enough modifications to the Hamiltonian
(\ref{H-orig}) so as to perform general quantum computing.

For a single qubit gate on qubit $k$ we can induce unitary transformations by
means of Stark shifts and transitions between internal states. During the whole
operation we set $J_{b,k}=0$ in order to isolate the qubits. The atom-laser
interaction is then described by the Hamiltonian
\begin{equation}
  \label{H1-real}
  H_{las,k} = \frac{\Delta_k}{2} (\ad_k a_k - \bd_k b_k) +
  \frac{\Omega_k}{2} (e^{i\varphi}\ad_k b_k + e^{-i\varphi}\bd_k a_k).
\end{equation}
For $U_{bb}\gg |\Delta_k|,|\Omega_k|$, we can replace (\ref{H1-real}) by an
effective Hamiltonian of the form
\begin{equation}
  \label{H1} H_1 = \frac{\Delta}{2} \sigma_z +
  \frac{\Omega}{2}(\sigma_+ e^{i\varphi} + \sigma_- e^{-i\varphi}),
\end{equation}
with $\Delta=\Delta_k$ and $\Omega = \Omega_k\sqrt{n_k-1}$.

For the realization of two-qubit operations we tilt the lattice using an
electric field, $H_{tilt} = \sum_k k g(\ad_ka_k + \bd_kb_k)$. The tilting must
be weak as to only virtual hopping of atoms of type $\ket{b}$ ($J_b \ll
|U_{bb}-g|$). After adiabatic elimination we find that the effective
Hamiltonian becomes
\begin{equation}
  \label{H2}
  H_2 = \tilde\Delta \ket{11}\bra{11},\quad
  \tilde\Delta = \frac{J_b^2}{g-U_{bb}}.
\end{equation}

The Hamiltonians $H_1$ and $H_2$ pose now two problems. The first one is that
$H_1$ depends on the occupation numbers. A traditional approach to quantum
computing would be to tune the parameters $\Delta$ and $\Omega$ and let the
resulting Hamiltonian operate for a time $T$, $U_{real} = \exp(-iH_1T)$.
However, since the parameters are unknown, we cannot take this na\"{\i}ve
route. The second difficulty resides on the magnitude of $J_b$, $g$, and
$U_{bb}$: these values are very sensitive to the properties of the lattice and
difficult to control. At most, we will be able to assure that $J_b$ and $g$ are
zero, or that $g$ is similar to $U_{bb}$; but we will be unable to fix the
value of $\tilde\Delta$ with enough accuracy that $U_{real}=\exp(-iH_2T)$
resembles a controlled-Z gate.

In the following section we will solve these problems. In \S\ref{sec:errors}
develop an abstract protocol which, up from the Hamiltonians (\ref{H1}) and
(\ref{H2}), produces a universal set of gates that can be used for quantum
computing. Next in \S\ref{sec:errors} we will study the influence of all the
processes which we have neglected in the abstract derivation, such as
interaction and hopping of atoms in state $\ket{a}$, sensitivity to occupation
numbers, etc.

\section{Computation with unknown parameters}
\label{sec:adiabatic}

\subsection{Basic ideas}

Let us consider a set of qubits that can be manipulated according to the single
qubit Hamiltonian (\ref{H1}) and the two--qubit Hamiltonian (\ref{H2}).  We
will assume that most of the parameters appearing in these Hamiltonians are
basically unknown. On the other hand, we will not consider any randomness in
these parameters because the corresponding errors may be corrected with
standard error correction methods (M. A. Nielsen 2002), as long as they are
small, and in most cases random quick fluctuations of the parameters $I$ will
be averaged out in the adiabatic process.

In particular we will assume that only the phase of the laser, $\varphi$, can
be precisely controlled.  For the other parameters we will impose that: (i)
they are given by an unknown (single valued) function of some experimentally
controllable parameters, (ii) they can be set to zero, and (iii) they are
positive\footnote{This is just accommodate the physical restrictions of
  \S\ref{sec:lattice}. The scheme actually becomes simpler when $\Delta$ or
  $\tilde\Delta$ may take negative values.}. For example, we may have
$\Omega=f(I)$, where $I$ is a parameter that can be experimentally controlled,
and we only know about $f$ that $f(0)=0$ and that we can reach some value
$\Omega_m\equiv f(I_m) \neq 0$ for some $I_m$\footnote{Note that, in many
  realistic implementations it is not possible to measure the dependence of
  these parameters (function $f$) because measurements are destructive (lead to
  heating or atom losses).  Therefore $f(I)$ is different in different
  experimental realizations.}.  Outside this, $f(I)$ may change in different
experimental realizations.

The physical scenario described in \S\ref{sec:lattice} corresponds to
this situation, but we want to stress that these conditions can be naturally
met in more general scenarios.  For example, the qubit states $\ket{0}$ and
$\ket{1}$ may correspond to two degenerate atomic (ground state) levels which
are driven by two lasers of the same frequency and different polarization. The
corresponding Hamiltonian is given by (\ref{H1}), where the parameters
$\varphi,\Omega,\Delta$ describe the relative phase of the lasers, the Rabi
frequency and detuning of the two-photon Raman transition, respectively. The
Rabi frequency can be changed by adjusting the intensity of the lasers, and the
detuning and the phase difference by using appropriate modulators.  In
practice, $\Omega$ ($\Delta$) can be set to zero very precisely by switching
off the lasers (modulators) and $\varphi$ may be very precisely controlled to
any number between 0 and $2\pi$. However, fixing $\Omega$ or $\Delta$ to a
precise value can be much more difficult.

\begin{figure}[t]
  \resizebox{\linewidth}{!}{\includegraphics{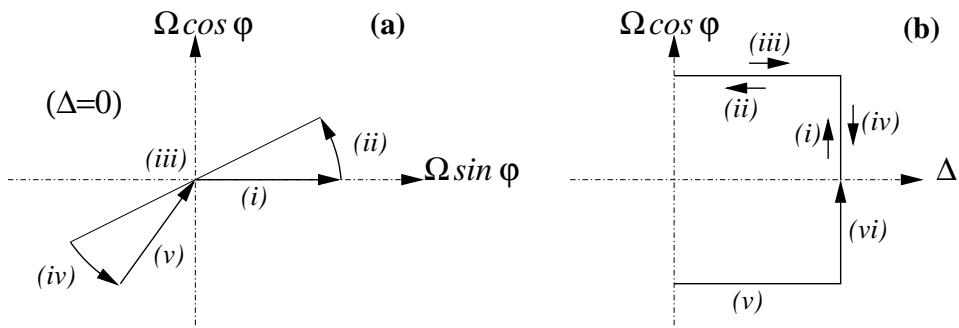}}
  \caption{
    \label{fig-local}
    Schema of how the parameters of Hamiltonian (\ref{H1}) have to be changed
    in order to perform a phase gate (a), and Hadamard gate (b).}
\end{figure}

The idea of obtaining perfect gates with unknown parameters combines the
techniques of adiabatic passage (P. Zanardi 1999; J. Pachos 2000) with ideas of
quantum control (L. Viola 1998; L.-M. Duan 1998). Let us briefly recall the
adiabatic theorem, which is a fundamental tool in our method.  Suppose we
have a Hamiltonian that depends parametrically on a set of parameters, denoted
by $p$, which are changed adiabatically with time along a given trajectory
$p(t)$.  After a time $T$, the unitary operator corresponding to the evolution
is
\begin{equation}
U(T)= \sum_\alpha e^{i(\phi_\alpha + \psi_\alpha)}
\ket{\Phi_\alpha[p(T)]}\bra{\Phi_\alpha[p(0)]}.
\end{equation}
Here, $|\Phi_\alpha(p)\rangle$ are the eigenstates of the Hamiltonian for which
the parameters take on the values $p$. The phase $\phi_\alpha$ is a dynamical
phase that explicitly depends on how the parameters $p$ are changed with time,
whereas the phase $\psi_\alpha$ is a purely geometrical phase ands depends on
the trajectory described in the parameter space.  Our basic idea to perform any
given gate is first to design the change of the parameters in the Hamiltonians
(\ref{H1})-(\ref{H2}) such that the eigenvectors evolve according to the
desired gate, and then to repeat the procedure changing the parameters
appropriately in order to cancel the geometric and dynamical phases.

\subsection{Local gates}

Using the previous ideas we are able to implement a universal set of gates,
which is made of a phase gate, $U=e^{i\theta\sigma_z/2}$, a Hadamard gate and a
CNOT gate. To perform the phase gate $U=e^{i\theta\sigma_z/2}$ we work with the
single-qubit Hamiltonian (\ref{H1}). We set $\Delta=0$ for all times and change
the remaining parameters $(\Omega,\varphi)$ as depictured in figure
\ref{fig-local}(a):
\begin{eqnarray}
  \label{protocol-phase}
  (0,0) &\stackrel{(i)}{\rightarrow}& (\Omega_m,0)
  \stackrel{(ii)}{\rightarrow} (\Omega_m,\theta/2)
  \stackrel{(iii)}{\Rightarrow} (\Omega_m,\theta/2+\pi) \nonumber \\
  &\stackrel{(iv)}{\rightarrow}& (\Omega_m,\theta+\pi)
  \stackrel{(v)}{\rightarrow} (0,\theta+\pi)
\end{eqnarray}
All steps are performed adiabatically and require a total time $T$, except for
step (iii) whose double arrow indicates a sudden change of parameters.  Note
that $\Omega(0)=\Omega(2T)=0$, $\Omega(t)=\Omega(2T-t)$ and $\varphi(t) = \pi +
\theta - \varphi(2T-t)$, which does not require the knowledge of the function
$f$ but implies a precise control of the phase.  A simple analysis shows that
(i-v) achieve the desired transformation $\ket{0} \to e^{i\theta/2} \ket{0}$,
$\ket{1} \to e^{-i\theta/2} \ket{1}$.  Note also that the dynamical and
geometrical phases acquired in the adiabatic processes (i-v) cancel out.

The Hadamard gate can be performed in a similar way. In the space
$[\Delta,\Omega_x=\Omega\cos(\varphi)]$, the protocol is
\begin{eqnarray}
  \label{protocol-hadamard}
  (0,\Omega_m) &\stackrel{(i)}{\rightarrow}&
  (\Delta_m,\Omega_m) \stackrel{(ii)}{\rightarrow}
  (\Delta_m,0) \stackrel{(iii)}{\rightarrow}
  (\Delta_m,\Omega_m) \nonumber\\
  &\stackrel{(iv)}{\rightarrow}&
  (0,\Omega_m) \stackrel{(v)}{\Rightarrow}
  (0,-\Omega_m) \stackrel{(vi)}{\rightarrow}
  (\Delta,-\Omega_m) \stackrel{(vii)}{\rightarrow}
  (\Delta,0),
\end{eqnarray}
as shown in figure \ref{fig-local}(b-c).  In order to avoid the dynamical
phases, we have to make sure that steps (i-v) are run in half the time as
(vi-vii).  More precisely, if $t<T$, we must ensure that
$\Delta(t)=\Delta(T-t)$, $\Omega_x(t)=\Omega_x(T-t)$, $\Delta(T+t) =
\Delta(t/2)$ and $\Omega_x(T+t) = \Omega_x(t/2)$. With this requisite we get
$\frac{1}{\sqrt{2}}(\ket{0} + \ket{1}) \to \ket{0}$,
$\frac{1}{\sqrt{2}}(\ket{0} - \ket{1}) \to -\ket{1}$.  Again, the whole
procedure does not require us to know $\Omega$ or $\Delta$, but rather to
control the evolution of the experimental parameters which determine them.

\subsection{Nonlocal gates}

The C-NOT gate requires the combination of two two-qubit processes using $H_2$
and one local gate. The first process involves changing the parameters
$[\tilde\Delta,\Omega_x=\Omega\cos(\varphi)]$ of equation (\ref{H2})
according to
\begin{eqnarray}
  \label{protocol-nonlocal}
  (\tilde\Delta_m,0) &\stackrel{(i)}{\rightarrow}&
  (\tilde\Delta_m,\Omega_m)
  \stackrel{(ii)}{\rightarrow} (0,\Omega_m)
  \stackrel{(iii)}{\Rightarrow} (0,-\Omega_m) \nonumber \\
  &\stackrel{(iv)}{\rightarrow}& (\tilde\Delta_m,-\Omega_m)
  \stackrel{(v)}{\rightarrow} (\tilde\Delta_m,0). \label{sigmay}
\end{eqnarray}
This procedure gives rise to the transformation
\begin{equation}
  U_1 = \ket{0}\bra{0}\otimes \one + e^{i\xi}\ket{1}\bra{1}\otimes i\sigma_y,
\end{equation}
where $\xi = \int_0^T \delta(t) dt$ is an unknown dynamical phase. The second
operation required is a NOT on the first qubit $U_2 = (\ket{0}\bra{1} +
\ket{1}\bra{0}) \otimes \one$.  Finally, if $\tilde\Delta^{(1)}(t)$ denotes the
evolution of $\tilde\Delta$ in equation  (\ref{sigmay}), we need to follow a path
such that $\tilde\Delta^{(3)}(t) = \tilde\Delta^{(1)}(t)$,
$\Omega^{(3)}(t) = 0$.  If the timing is correct, we achieve
$U_3 = (\ket{0}\bra{0} + e^{i\xi}\ket{1}\bra{1})\otimes \one$.
Everything combined gives us the CNOT up to a global unimportant phase
$U_{cnot} = \ket{0}\bra{0}\otimes \one +\ket{1}\bra{1}\otimes i\sigma_y
  = e^{-i\xi}U_2 U_3 U_2 U_1$.


\begin{figure}[t]
  \centering
  \resizebox{0.4\linewidth}{!}{\includegraphics{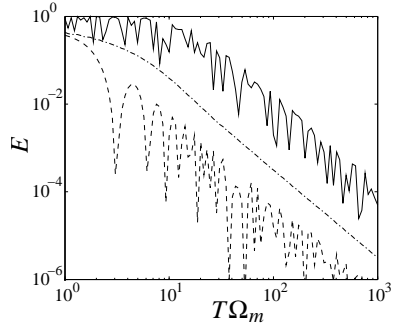}}
  \caption{\label{fig-error1}
    Log-log plot of the gate error, $E=1-{\cal F}$, for the Hadamard (solid),
    phase (dashed) and CNOT (dotted) gates, when we use the ideal Hamiltonians
    (\ref{H1}) and (\ref{H2}) to simulate them. The parameters for the
    simulations are $\{\Delta_m/\Omega_m = \tilde\Delta_m/\Omega_m = 1/10,
    \varphi_m=\pi/4\}$, and we change the speed of the process to study the
    influence of non--adiabaticity.}
\end{figure}

\section{Control of errors}
\label{sec:errors}

In this section we study whether it is feasible to apply the methods developed
in \S\ref{sec:adiabatic} to the physical setup envisioned in
\S\ref{sec:lattice}. First of all we will study how fast the operations from
\S\ref{sec:adiabatic} must be performed in order to minimize the deviations
from the adiabatic theorem. And second and most important, we have to consider
contributions to the energy which escape the terms considered in equations
(\ref{H1}) and (\ref{H2}). We will analyse both sources of error separately,
combining analytical estimates with numerical simulations of our techniques for
small number of atoms.

\subsection{Adiabaticity}

To study the sensitivity of our method against non--adiabatic processes, we
have simulated the protocols (\ref{protocol-phase})-(\ref{protocol-nonlocal})
using the ideal model given by Hamiltonians (\ref{H1}) and (\ref{H2}). For each
of the gates we have fixed all parameters $\{\Delta_m/\Omega_m =
\tilde\Delta_m/\Omega_m = 1/10, \varphi_m=\pi/4\}$ except the time, and then we
have computed how the error decreases as we decrease the speed of the adiabatic
passage. The results are shown in figure \ref{fig-error1}. As a figure of merit
we have chosen the gate fidelity (M. A. Nielsen 2002)
\begin{equation}
{\cal F} = 2^{-n}|\mathrm{Tr}\{U_{ideal}^\dagger U_{real}\}|^2
\end{equation}
where $n$ is the number of qubits involved in the gate, $U_{ideal}$ is the gate
that we wish to produce and $U_{real}$ is the actual operation performed.  As
expected, the adiabatic theorem applies when the processes are performed with a
sufficiently slow speed.  Typically a time $T\sim 300/\Omega_m$ is required for
the desired fidelity ${\cal F}=1-10^{-4}$.

It is also worth mentioning, that in figure \ref{fig-error1} and
\ref{fig-error2}, strong, rapid oscillations of the error are seen. These
oscillations are due to either the nonadiabaticity of the process (figure
\ref{fig-error1}), or to imperfections in the Hamiltonian (figure
\ref{fig-error2}). For a two-level system undergoing adiabatic evolution it is
easy to prove that, while the amplitude of the oscillations is proportional to
the speed of the adiabatic process, the frequency is instead related to the
energy difference between contiguous eigenspaces. This frequency is for us
unknown, and consequently, these oscillations may not be used to improve the
accuracy of our method by looking for some `magic times'.

\subsection{Imperfections}

Outside the non--adiabaticity of a real experiment, there are two other sources
of error which we must consider. (i) The quotient $J_b/U_{bb}$ is nonzero,
which means that more than one atom per well may be excited. (ii) Atoms in
state $\ket{a}$ interact and hop. This introduces new terms in equation
(\ref{H-orig}), which are of the form $J_{a,k}(\ad_ka_{k+1}+\ad_{k+1}a_k)$,
$U_{aa}\ad_k\ad_ka_ka_k$, and $U_{ab}\ad_k\bd_ka_kb_k$. And finally, (iii)
atoms in either state may jump to neighbouring sites, permanently changing the
occupation numbers.

\begin{figure}[t]
  \centering
  \resizebox{\linewidth}{!}{\includegraphics{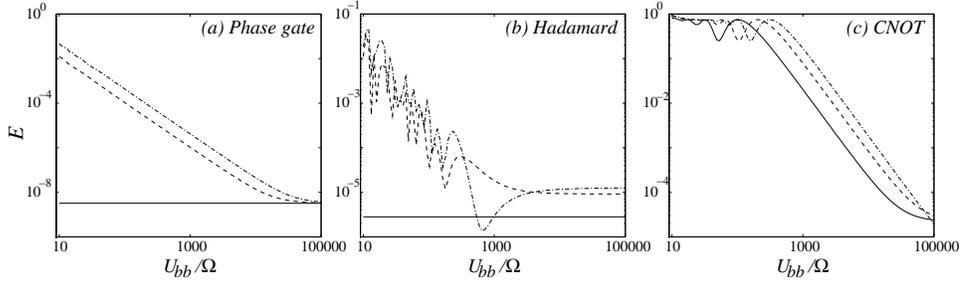}}
  \caption{\label{fig-error2}
    Log-log plot of the gate error, $E=1-{\cal F}$, for the (a) phase, (b)
    Hadamard and (c) CNOT gates. The parameters for the simulations are
    $U_{aa}=U_{ab}=J_m$, $J_a=J_b$,$g=U_{bb}+U_{ab}/2$, and
    $T=100/\Omega_m$. For the local gates we choose $\Delta_m=6\Omega_m=1$,
    and for the nonlocal gate $\Omega_m=J_m^2/6$. For each simulation we
    choose different population imbalance ($|n-m|=0,1,2$ for solid, dashed
    and dotted lines), and change the interaction constant $U_{bb}$.}
\end{figure}

The effects (i) and (iii) are suppressed if the coupling between internal
levels and the amplitude of hopping are both small compared to the energy gap
between our computation space and the undesired excitations. In other words,
we require
\begin{equation}
  (\Omega/U_{bb})^2 \ll 1\quad \mathrm{and}\quad (J_{a,k}/U_{bb})^2\ll 1.
\end{equation}
To analyse the remaining errors we develop an effective Hamiltonian which
contains (\ref{H1-real}) and (\ref{H-orig}) plus new terms
($J_a,U_{ab},U_{aa}$) that we did not consider before. In equation
(\ref{H1-real}), the virtual excitation of two atoms increments the parameter
$\Delta$ by an unknown amount, $\Delta_{\mathrm{eff}} \sim \Delta+ 2\Omega^2n_k
/ (\Delta+U_{ab}-U_{bb})$. If
\begin{equation}
U_{ab}\ll U_{bb}\quad\mathrm{and}\quad  \Omega^2n_k T/U_{bb} \ll 1,
\end{equation}
this shift may be neglected.  In the two-qubit gates the energy shifts are
instead due to virtual hopping of all types of atoms, and they are also
accompanied by the possibility of swapping both qubits ($\ket{01}
\leftrightarrow \ket{10}$). Both contributions are of the order of
$\max(J_b,J_a)^2/g^2 \sim J^2/U_{bb}$, and for
\begin{equation}
J^2T/U_{bb} \ll 1  
\end{equation}
they also may be neglected.

To quantitatively determine the influence of these errors we have simulated the
evolution of two atomic ensembles with an effective Hamiltonian which results
of applying second order perturbation theory to equation (\ref{H-orig}), and
which takes into account all important processes. The results are shown in
figures \ref{fig-error2}. In these pictures we show the error of the gates for
simulations in which all parameters are fixed, except for $U_{bb}$ and the
occupation numbers of the wells. The first conclusion is that the stronger the
interaction between atoms in state $\ket{b}$, the smaller the energy shifts.
This was already evident from our analytical estimates, because all errors are
proportional to $1/U_{bb}$. Typically, a ratio $U_{bb} = 10^4 U_{ab}$ is
required to make ${\cal F} = 1-10^{-4}$.  Second, the larger the number of
atoms per well, the poorer the fidelity of the local gates [Figure
\ref{fig-error2}(a-b)]. And finally, as figure \ref{fig-error2}(c) shows, the
population imbalance between wells influences very little the fidelity of the
two-qubit gate.

\section{Conclusions}
\label{sec:conclusions}

In this work we have shown that it is possible to perform quantum computation
with cold atoms in a tunable optical lattice. Our scheme is based on performing
adiabatic passage with one-qubit (\ref{H1}) and two-qubit (\ref{H2})
Hamiltonians. With selected paths and appropriate timing, it is possible to
perform a universal set of gates (Hadamard gate, phase gate and a CNOT).
Thanks to the adiabatic passage, the proposal works even when the number of
atoms per lattice is unknown or the constants in the governing Hamiltonians
have large uncertainties.  These procedures can not only be used for quantum
computing but also for quantum simulation (E. Jan\'e 2002), and the same ideas
can also be applied to other setups like the micro-traps demonstrated in R.
Dumke (2002).

\begin{acknowledgements}
We thank D. Liebfried and P. Zoller for discussions and the EU project EQUIP
(contract IST-1999-11053).
\end{acknowledgements}

\label{lastpage}
\end{document}